\documentclass[epj]{svjour}
\usepackage{graphics}
\usepackage{epstopdf}
\usepackage{amsmath}
\usepackage{marvosym}
\usepackage{color}
\UseRawInputEncoding

\begin{document}
\title{A verifiable quantum key agreement protocol based on six-qubit cluster states}
\author{Li-Juan Liu\inst{1} \and Zhi-Hui Li\inst{1}
\thanks{\emph{e-mail:lizhihui@suun.edu.cn} }
}                     
%
%
\institute{$^{1}$ School of Mathematics and Statistics, Shaanxi Normal University, ${\rm Xi^{'}}$an 710119, P.R. China}
\date{Received: date / Revised version: date}
%
\abstract{
Quantum key agreement requires all participants to recover the shared key together, so it is crucial to resist the participant attack. In this paper, we propose a verifiable multi-party quantum key agreement protocol based on the six-qubit cluster states. A verifiable distributor who preserves some subsequences of the six-qubit cluster states is introduced into this protocol, thus the participants can not obtain the shared key in advance. Besides, the correctness and simultaneity of the shared key are guaranteed by the trusted design combiner and homomorphic hash function. Furthermore, the security analysis shows that the new protocol can resist the external and internal attacks.
\PACS{
      {}{}   \and
      {}{}
     } 
} 
\maketitle
\section{Introduction}
Quantum cryptography is an interdiscipline that combines classical cryptography with quantum mechanics. Different from some cryptosystems based on mathematical complexity, its theoretical basis is quantum mechanics, namely uncertainty principle and non-cloning principle. Therefore, it has overwhelming advantages in the information security. Nowadays, with the rapid development of the quantum cryptography\cite{refA}, the research of it is extremely active in the fields of quantum key distribution (QKD) \cite{refB,refC,refD}, quantum secret sharing (QSS)\cite{refE,refF,refG} and quantum secure direct communication (QSDC) \cite{refH,refI}. Unlike the QKD protocols, quantum key agreement (QKA) \cite{refJ,refK,refL,refM,refN} is a new important branch that each participant plays an equivalent role on generating the shared key.

In 2004, Zhou et al. \cite{refO} proposed a groundbreaking QKA scheme by using quantum teleportation technique, in which two participants can generate a shared key over public channels. However, Tsai et al. \cite{refP} found that it cannot resist the participant attacks. A QKA scheme based on maximally entangled states was put forward by Hsueh and Chen\cite{refQ} in the same year, but it was discovered that only one participant can recover the shared key. In 2010, a two-party QKA scheme which uses the delayed measurement method and the standard BB84 protocol \cite{refR} was proposed by Chong and Hwang \cite{refS}. And Shen et al. \cite{refT} presented a two-party QKA scheme based on four-qubit cluster states in 2014. Unfortunately, it only applied to two-party, and it's obvious that multi-party quantum key agreement (MPQKA) schemes are more applicable to the actual situation.  Thus, Shi and Zhong \cite{refU} extended from two-party QKA protocols to multi-party successfully, and proposed the first MPQKA scheme. Subsequently, many MPQKA schemes \cite{refV,refW,refX,refY} were proposed.

 Security is crucial in the quantum cryptography, so the participant attacks \cite{refZ,refa} should be paid more attention in the multi-party
 environment. In 2019, Liu et al. \cite{refb} proposed a high efficient MPQKA protocol by using the four-qubit cluster states
 ingeniously. However, we find that it could not resist the participant attacks effectively. Because all participants can't know the value of
 the shared key and can't recover the shared key at the same time, the internal fraudster interferes with the other participants after
 recovering the shared key which will lead to an error shared key but cannot be found.

In this paper, we propose a new MPQKA protocol based on the six-qubit cluster states. It can not only realize the recovery of the shared key by all participants, but also guarantee the correctness and simultaneity of the recovered shared key. The advantages of the protocol are as follows:

\begin{enumerate}
\item[1)]A distributor whose each operation can be verified is introduced. The distributor distributes the remaining subsequences after preserving two subsequences. Therefore, the shared key will not be recovered in advance by the internal fraudsters or external eavesdroppers during the transmission process.
\item[2)]In the process of transmitting the information by the subsequences, the decoy particles are always inserted randomly to resist the eavesdropping attacks. Therefore, it can avoid eavesdropping by encrypting transmission.
\item[3)] The digital authentication can be achieved through the following steps. Firstly, the six-qubit cluster states prepared by the distributor can be verified by the participants through the related properties of six-qubit cluster states. Secondly, before recovering the shared key, the trusted design combiner (TDC) \cite{refF} will use the homomorphic hash function \cite{refc} to detect the received information. Both of them can prevent information tampering, so as to effectively ensure the realization of digital authentication.
\item[4)]The shared key is recovered by all participants, and the correctness and simultaneous of it can be guaranteed. Besides, the participant attacks can be resisted successfully.
\end{enumerate}

The structure of this paper is as follows. This basic knowledge is introduced in Section 2, including the related comments of six-qubit cluster states, the homomorphic hash function and the trusted design combiner. Models for verifiable quantum key agreement is presented in section 3. A verifiable MPQKA protocol with six-qubit cluster states is put forward in Section 4. The security of this protocol is discussed in Section 5, which includes the external and internal attacks. In Section 6, the performance analysis of this protocol is discussed, the correctness analysis and comparation analysis are included. In Section 7, namely the last part, this paper is summarized.

\section{Basic knowledge}
Next, the basic knowledge used in the protocol design is introduced, including the the X operation and six-qubit cluster states, related properties of six-qubit cluster states, the homomorphic hash function and the trusted design combiner.
\subsection{the X operation and six-qubit cluster states}
Now, we will introduce the X operation \cite{refb} used in our protocol: $X=|0\rangle\langle1|+|1\rangle\langle0|$. This operation can realize the flip of the particles, i.e., $X|0\rangle = |1\rangle$, $X|1\rangle = |0\rangle$.

Then we use six-qubit cluster states as quantum resource, that is,
\[
|C\rangle=\frac{1}{2}\left(|000000\rangle+|000111\rangle+|111000\rangle-|111111\rangle\right)_{123456}.
\]

Assuming that Bob and David generate their secret keys randomly:
\[
K_{B}=(K_{B}^{\left(1\right)},K_{B}^{\left(2\right)},\cdots,K_{B}^{\left(m\right)}),
\]
\[
K_{D}=(K_{D}^{\left(1\right)},K_{D}^{\left(2\right)},\cdots,K_{D}^{\left(m\right)}).
\]
where $K_{B}^{\left(j\right)},K_{D}^{\left(j\right)}\in\{00,01,10,11\}$, $j=1,2,\cdots,m$. Bob performs the X operation on the particles 3 and 4 of $|C\rangle$ according to $K_{B}$. That is, if the first bit of $K_{B}$ is 0 (1), the particle 3 is stationary (flipped). If the second bit of $K_{B}$ is 0 (1), the particle 4 is stationary (flipped). And David performs the X operation on the particles 5 and 6 of $|C\rangle$ according to $K_{D}$. If the first bit of $K_{D}$ is 0 (1), the particle 5 is stationary (flipped). If the second bit of $K_{D}$ is 0 (1), the particle 6 is stationary (flipped). Therefore, we will obtain one of the following 16 cluster states:
\[
|C_{1}\rangle=\frac{1}{2}(|000000\rangle+|000111\rangle+|111000\rangle-|111111\rangle)_{123456},
\]
\[
|C_{2}\rangle=\frac{1}{2}(|000001\rangle+|000110\rangle+|111001\rangle-|111110\rangle)_{123456},
\]
\[
|C_{3}\rangle=\frac{1}{2}(|000010\rangle+|000101\rangle+|111010\rangle-|111101\rangle)_{123456},
\]
\[
|C_{4}\rangle=\frac{1}{2}(|000011\rangle+|000100\rangle+|111011\rangle-|111100\rangle)_{123456},
\]
\[
|C_{5}\rangle=\frac{1}{2}(|000100\rangle+|000011\rangle+|111100\rangle-|111011\rangle)_{123456},
\]
\[
|C_{6}\rangle=\frac{1}{2}(|000101\rangle+|000010\rangle+|111101\rangle-|111010\rangle)_{123456},
\]
\[
|C_{7}\rangle=\frac{1}{2}(|000110\rangle+|000001\rangle+|111110\rangle-|111001\rangle)_{123456},
\]
\[
|C_{8}\rangle=\frac{1}{2}(|000111\rangle+|000000\rangle+|111111\rangle-|111000\rangle)_{123456},
\]
\[
|C_{9}\rangle=\frac{1}{2}(|001000\rangle+|001111\rangle+|110000\rangle-|110111\rangle)_{123456},
\]
\[
|C_{10}\rangle=\frac{1}{2}(|001001\rangle+|001110\rangle+|110001\rangle-|110110\rangle)_{123456},
\]
\[
|C_{11}\rangle=\frac{1}{2}(|001010\rangle+|001101\rangle+|110010\rangle-|110101\rangle)_{123456},
\]
\[
|C_{12}\rangle=\frac{1}{2}(|001011\rangle+|001100\rangle+|110011\rangle-|110100\rangle)_{123456},
\]
\[
|C_{13}\rangle=\frac{1}{2}(|001100\rangle+|001011\rangle+|110100\rangle-|110011\rangle)_{123456},
\]
\[
|C_{14}\rangle=\frac{1}{2}(|001101\rangle+|001010\rangle+|110101\rangle-|110010\rangle)_{123456},
\]
\[
|C_{15}\rangle=\frac{1}{2}(|001110\rangle+|001001\rangle+|110110\rangle-|110001\rangle)_{123456},
\]
\[
|C_{16}\rangle=\frac{1}{2}(|001111\rangle+|001000\rangle+|110111\rangle-|110000\rangle)_{123456}.
\]

We can find the corresponding relationship between the secret key and the transformed cluster states in the Table 1. For the sake of consistency, the usage of $K_ {B}$ and $K_ {D}$ is always as above in this article.

\begin{table}[h]
\centering
\footnotesize
\caption{\footnotesize The relationship between the secret key and the transformed cluster states}


\begin{tabular}{|c|c|c|}
\hline
$j$-th two bits of $K_{B}$&$j$-th two bits of $K_{D}$&Final state\\
\hline
00 & 00 & $|C_{1}\rangle$\\
00 & 01 & $|C_{2}\rangle$\\
00 & 10 & $|C_{3}\rangle$\\
00 & 11 & $|C_{4}\rangle$\\
01 & 00 & $|C_{5}\rangle$\\
01 & 01 & $|C_{6}\rangle$\\
01 & 10 & $|C_{7}\rangle$\\
01 & 11 & $|C_{8}\rangle$\\
10 & 00 & $|C_{9}\rangle$\\
10 & 01 & $|C_{10}\rangle$\\
10 & 10 & $|C_{11}\rangle$\\
10 & 11 & $|C_{12}\rangle$\\
11 & 01 & $|C_{13}\rangle$\\
11 & 10 & $|C_{14}\rangle$\\
11 & 00 & $|C_{15}\rangle$\\
11 & 11 & $|C_{16}\rangle$\\
\hline
\end{tabular}
\end{table}


\subsection{Related properties of six-qubit cluster states}
\subsubsection{the properties of six-qubit cluster states under the X-basis}

The six-qubit cluster state $|C\rangle$ under the X-basis can be expressed as:
\begin{equation}
\begin{aligned}
|C\rangle&=\frac{1}{4}\left(|++++++\rangle+|+++--+\rangle\right)\nonumber\\
&+\frac{1}{4}\left(|++++--\rangle+|+++-+-\rangle\right)\nonumber\\
&+\frac{1}{4}\left(|--++++\rangle+|--+--+\rangle\right)\nonumber\\
&+\frac{1}{4}\left(|--++--\rangle+|--+-+-\rangle\right)\nonumber\\
&+\frac{1}{4}\left(|+--+++\rangle+|+----+\rangle\right)\nonumber\\
&+\frac{1}{4}\left(|+--+--\rangle+|+---+-\rangle\right)\nonumber\\
&+\frac{1}{4}\left(|-+-+++\rangle+|-+---+\rangle\right)\nonumber\\
&+\frac{1}{4}\left(|-+-+--\rangle+|-+--+-\rangle\right).\
\end{aligned}
\end{equation}

First of all, some specific stipulations are made. When the particles 1 and 2 of $|C\rangle$ are $|++\rangle$ or $|--\rangle$, namely the result of particles 1 and 2 are the same when they are measured with the X-basis, the measurement result is recorded as 0. When the particles 1 and 2 of $|C\rangle$ are $|+-\rangle$ or $|-+\rangle$, namely the result of particles 1 and 2 are different when they are measured with the X-basis, the measurement result is recorded as 1. Similarly, when the particles 3 and 4 (or the particles 5 and 6) are the same by measuring with the X-basis, it is recorded as 0. Otherwise, it is recorded as 1.

It can be seen from the above formula that the six-qubit cluster states $|C\rangle$ under the X-basis have the following properties:
\begin{enumerate}
\item[1)]When the measurement result of particles 1 and 2 is 0, the measurement results of particles 3, 4 and 5, 6 are either 0 simultaneously or 1 simultaneously.
\item[2)]When the measurement result of particles 1 and 2 is 1, the measurement results of particles 3, 4 and 5, 6 are inevitable to exist that one is 0 and the other is 1.
\end{enumerate}
\subsubsection{the properties of six-qubit cluster states under the Z-basis}
The six-qubit cluster state $|C\rangle$ under the Z-basis can be expressed as:
\[
|C\rangle=\frac{1}{2}\left(|000000\rangle+|000111\rangle+|111000\rangle-|111111\rangle\right)_{123456}.
\]

At first, we make some specific stipulations. When two particles are $|00\rangle$, namely the both results of two particles are $|0\rangle$ when they are measured with the Z-basis, the measurement result is recorded as $A$. Simliarly, when two particles are $|11\rangle$, the measurement result is recorded as $B$. When two particles are $|01\rangle$ or $|10\rangle$, the measurement result is recorded as $C$.

It can be seen from the above formula that the six-qubit cluster states $|C\rangle$ under the Z-basis have the following properties:
\begin{enumerate}
\item[1)]When the measurement result of particles 1 and 2 is $A$, the measurement results of particles 3, 4 and 5, 6 are either $A$ simultaneously or one is $B$ and the other is $C$.
\item[2)]When the measurement result of particles 1 and 2 is $B$, the measurement results of particles 3, 4 and 5, 6 are either $B$ simultaneously or one is $A$ and the other is $C$.
\end{enumerate}
\subsection{Homomorphic hash}
The hash function H of the additive homomorphism \cite{refc} has the following characteristics: all $x$ and $y$ satisfy the equation: $H(x)H(y)=H(x+y)$.

It should be noted that here the group operation in the output domain is arbitrary, but usually a product.

For instance, the hash function H: $Z_{p}\rightarrow Z_{l}$ can be constructed around the discrete-log assumption by a generator g as $H(v)=g^{v} \ {\rm mod} \ l$, which generates a collision-resistant hash as discussed in the literature\cite{refd}.

It is easy to verify the homomorphism by calculation:
\[
H(v_{1})H(v_{2})=g^{v_{1}}g^{v_{2}}=g^{v_{1}+v_{2}}=H(v_{1}+v_{2}) (all\ {\rm mod}\ l).
\]

It is discussed in the literature \cite{refe} about selecting a suitable generator.


\subsection{Trusted design combiner}
The trusted design combiner (TDC) is similar to an intelligent black box. It not only has the black box attributes, but also can handle some data intelligently, that is, simply calculate the obtained data.


\section{Models for verifiable quantum key agreement}
\subsection{System model}
The proposed protocol includes a distributor, $n$ ($n$ is an odd number) participants  ${\rm P}_{1},{\rm P}_{2},\cdots,{\rm P}_{n}$ and some internal and external adversaries. Assuming that all of them have unlimited computional power, and the identities of the participants are public, thus the identity authentication of the receiver is not required when transmitting information.

It is assumed that there is a secure channel between the distributor and every participant, so that the information can be transmitted to the receiver securely. In addition, assuming that the distributor and all participants are connected to a common authenticated
broadcast channel $\ell$, so that any message sent through $\ell$ can be heard by other receivers. An adversary cannot modify a message sent by an honest sender through $\ell$, or prevent an honest receiver from receiving a message from $\ell$
. Note that these assumptions have been widely used in existing quantum protocols. With these assumptions, we can focus our discussion on the key aspects of our protocol, rather than pay attention to the lower level of technical detail.

The protocol we proposed includes five phases. During the distributor operation phase, the distributor constructs six-qubit cluster states and obtains six subsequences respectively. In the case of retaining the first two subsequences, the remaining four subsequences are sent to two determined participants through the secure channel. The eavesdropping detection is realized by secure channel and $\ell$. When the two participants authenticate the identity of the distributor, they detect the correctness of the six-qubit cluster state prepared by the distributor through the secure channel. And in the participant operation phase, the participant sends the subsequence of X operation and realizes eavesdropping detection by using secure channel and $\ell$.

\subsection{Adversary model}
\subsubsection{Internal adversary}
The internal adversary is a legitimate internal distributor or participant. The internal adversary may carry out the forge attack alone, which make all participants get the wrong shared key, or collude with other internal adversaries to recover the shared key in advance. In our protocol, we assume that the distributor and the participants can not carry out joint attack, that is, the internal attack is divided into participant attack and distributor attack. At the same time, when the internal adversary obtains the information of the shared key, it will not disclose to the external adversary.
\subsubsection{External adversary}
The external adversary is an attacker who does not have any shared key information, but can try to obtain unauthorized access to the shared key information.

\section{Protocol description}
Based on six-qubit cluster states, we put forward a verifiable MPQKA in this section. It includes the distributor Alice and $n$ participants ${\rm P}_{1},{\rm P}_{2},\cdots,{\rm P}_{n}$, where $n$ is an odd number. And  each participant has the only public identity $x_{j}(j=1,2,\cdots,n)$.

Each participant ${\rm P}_{j} (j=1,2,\cdots,n)$ randomly generates a $2m$-bit string
\[
K_{j}=(K_{j}^{\left(1\right)},K_{j}^{\left(2\right)},\cdots,K_{j}^{\left(m\right)}),
\]
as his secret key. And he calculates the corresponding hash function value $H(j)$ with identity by $K_{j}$, where H is the homomorphic hash function. Then he will send $H(j)$ to the TDC.

All participants negotiate and publish the participant identity $x_{i}$ which is needed during the process of protocol.

\subsection{Distributor preparation}
\begin{enumerate}
\item[1)]Alice prepares $m+l$ six-qubit cluster states $|C\rangle$, where $m$ of $|C\rangle$ are used for participants to transfer information to generate shared key, and the remaining $l$ of them can verify the correctness of the six-qubit cluster states prepared by Alice.
Then she picks up the $k$-th qubit from each six-qubit cluster state to compose the subsequence $S_{k}$, where $k=1,2,3,4,5,6$. Alice keeps $S_{1}$ and $S_{2}$ for herself, then randomly selects enough decoy particles from four quantum states $\{|0\rangle,|1\rangle,|+\rangle,|-\rangle\}$, and randomly inserts them to obtain subsequences $S_{k}^{*} (k=3,4,5,6)$. At last, Alice sends $S_{3}^{*}$, $S_{4}^{*}$ to ${\rm P}_{i-1}$, and sends $S_{5}^{*}$, $S_{6}^{*}$ to ${\rm P}_{i+1}$, where $i$ is random, and $i=1,2,\cdots,n$.
\item[2)]After confirming that ${\rm P}_{i-1}$ $({\rm P}_{i+1})$ has received $S_{3}^{*}$ and $S_{4}^{*}$ ($S_{5}^{*}$ and $S_{6}^{*}$), Alice announces the positions of the decoy particles as well as the corresponding measurement bases. Then ${\rm P}_{i-1}$ $({\rm P}_{i+1})$ uses the given measurement bases to measure the decoy particles and informs Alice about the measurement results. Finally, Alice calculates the error rate according to the initial states of the decoy particles and the measurement results. If the error rate is less than predetermined value, ${\rm P}_{i-1}$ $({\rm P}_{i+1})$ will recover $S_{3}$ and $S_{4}$ ($S_{5}$ and $S_{6}$), and proceed to the next step. Otherwise, she will abandon this protocol and prepare the new subsequences.
\end{enumerate}

\subsection{Authentication of distributor identity}
${\rm P}_{i-1}$ and ${\rm P}_{i+1}$ randomly specify the positions of $l$ particles from $\{1,2,\cdots,m+l\}$ and stipulate to uses X-basis or Z-basis. They require Alice to measure the corresponding positions of $S_{1}$ and $S_{2}$ with specified basis and announce the measurement result. When they receive the measurement result, ${\rm P}_{i-1}$ and ${\rm P}_{i+1}$ will measure the particles of the corresponding positions of $S_{3}$, $S_{4}$ and $S_{5}$, $S_{6}$ with the specified basis.

\begin{enumerate}
\item[1)]If the specified basis is X-basis, when Alice announces the measurement result is 0, the measurement results of ${\rm P}_{i-1}$ and ${\rm P}_{i+1}$ are either 0 or 1 simultaneously. When she announces the measurement result is 1, the measurement results of ${\rm P}_{i-1}$ and ${\rm P}_{i+1}$ must satisfy that one is 0 and the other is 1. Then it can be confirmed that Alice prepares   $|C\rangle$ correctly through the properties of six-qubit cluster states under the X-basis.
\item[2)]If the specified basis is Z-basis, when Alice announces the measurement result is $A$, the measurement results of ${\rm P}_{i-1}$ and ${\rm P}_{i+1}$ are either $A$ simultaneously or one is $B$ and the other is $C$. When she announces the measurement result is $B$, the measurement results of ${\rm P}_{i-1}$ and ${\rm P}_{i+1}$ must be either $B$ simultaneously or one be $A$ and the other be $C$. Then it can be confirmed that Alice prepares $|C\rangle$ correctly through the properties of six-qubit cluster states under the Z-basis.
\end{enumerate}

If Alice's identity is correct, ${\rm P}_{i-1}$ and ${\rm P}_{i+1}$  will eliminate $l$  particles of the corresponding positions from $S_{k}$, and they will form new subsequences  $S_{k}^{'}$, where  $k=1,2,3,4,5,6$. Then the next step is proceeded to.

\subsection{Participants operation}
\begin{enumerate}
\item[1)]According to $K_{i-1}^{\left(j\right)}$ ($K_{i+1}^{\left(j\right)}$)$(i=1,2,\cdots,n;j=1,2,\cdots,$

$m)$, ${\rm P}_{i-1}$ $({\rm P}_{i+1})$  performs the X operation on the  $j$-th of $S_{3}^{'}$  and the  $j$-th of  $S_{4}^{'}$ (the  $j$-th of $S_{5}^{'}$  and the  $j$-th of $S_{6}^{'}$). Then he can get $S_{3}^{\left(i-1\right)}$ and $S_{4}^{\left (i-1\right)}$ ($S_{5}^{\left(i+1\right)}$ and $S_{6}^{\left(i+1\right)}$).
This rule is described as follows. If the first bit of $K_{i-1}^{\left(j\right)}$  is 0 (1), the  $j$-th position of $S_{3}^{'}$  is stationary (flipped). If the second bit of $K_{i-1}^{\left(j\right)}$  is 0 (1), the  $j$-th position of $S_{4}^{'}$  is stationary (flipped). If the first bit of  $K_{i+1}^{\left(j\right)}$ is 0 (1), the  $j$-th position of $S_{5}^{'}$  is stationary (flipped). If the second bit of $K_{i+1}^{\left(j\right)}$  is 0 (1), the  $j$-th position of  $S_{6}^{'}$ is stationary (flipped).
\item[2)]${\rm P}_{i-1}$ $({\rm P}_{i+1})$ randomly selects enough decoy particles and inserts them into  $S_{3}^{\left(i-1\right)}$ and $S_{4}^{\left (i-1\right)}$ ($S_{5}^{\left(i+1\right)}$ and $S_{6}^{\left(i+1\right)}$), and obtains $S_{3}^{\left(i-1\right)*}$ and $S_{4}^{\left (i-1\right)*}$ ($S_{5}^{\left(i+1\right)*}$ and $S_{6}^{\left(i+1\right)*}$). Then ${\rm P}_{i-1}$ $({\rm P}_{i+1})$  will send  $S_{3}^{\left(i-1\right)*}$ and $S_{4}^{\left(i-1\right)*}$ ($S_{5}^{\left(i+1\right)*}$ and $S_{6}^{\left(i+1\right)*}$) to ${\rm P}_{i-2}$ $({\rm P}_{i+2})$.
\item[3)]As ${\rm P}_{i-1}$  and ${\rm P}_{i+1}$ did in the step 1) and 2), the participants  ${\rm P}_{i-2},{\rm P}_{i+2}$, $ \cdots, {\rm P}_{i-\frac{n-3}{2}}$  and ${\rm P}_{i+\frac{n-3}{2}}$  perform the eavesdropping detections and the X operation. This process will be terminated until ${\rm P}_{i-\frac{n-1}{2}}$(${\rm P}_{i+\frac{n-1}{2}}$) obtains $S_{3}^{\left(i-\frac{n-1}{2}\right)*}$ and $S_{4}^{\left(i-\frac{n-1}{2}\right)*}$($S_{5}^{\left(i+\frac{n-1}{2}\right)*}$ and

    $S_{6}^{\left(i+\frac{n-1}{2}\right)*}$) and sends them to Alice.
\end{enumerate}

\subsection{Measurement}
By the same method as the step 2) of 4.1, Alice can recover  $S_{3}^{\left(i-\frac{n-1}{2}\right)}$, $S_{4}^{\left(i-\frac{n-1}{2}\right)}$, $S_{5}^{\left(i+\frac{n-1}{2}\right)}$ and $S_{6}^{\left(i+\frac{n-1}{2}\right)}$  without exceeding predetermined value of the error rate. Then Alice combines the  $j$-th particle of $S_{1}^{'}$, $S_{2}^{'}$, $S_{3}^{\left(i-\frac{n-1}{2}\right)}$, $S_{4}^{\left(i-\frac{n-1}{2}\right)}$, $S_{5}^{\left(i+\frac{n-1}{2}\right)}$ and $S_{6}^{\left(i+\frac{n-1}{2}\right)}$, and measures them with the cluster basis respectively, where $j=1,2,\cdots,m$. Thus, Alice can get   $K_{B}$ and  $K_{D}$ through the Table~\ref{Table:1}, then calculate the corresponding hash function values $H(B)$  and  $H(D)$.

\subsection{TDC operation}
Alice sends $K_{B}$ and  $K_{D}$,  $H(B)$ and  $H(D)$ to the TDC.  Then the $i$-th participant ${\rm P}_{i}$ sends the secret key $K_{i}$ to the TDC.
\begin{enumerate}
\item[1)]The following verifications will be performed in the TDC:
\begin{equation}
\label{eq1}
H(B)=H(i-\frac{n-1}{2})H(i-\frac{n-3}{2})\cdots H(i-1).
\end{equation}
\begin{equation}
\label{eq2}
H(D)=H(i+1)H(i+2)\cdots H(i+\frac{n-1}{2}).
\end{equation}
\begin{equation}
\label{eq3}
H(B)=H(K_{B}).
\end{equation}
\begin{equation}
\label{eq4}
H(D)=H(K_{D}).
\end{equation}
\begin{equation}
\label{eq5}
H(i)=H(K_{i}).
\end{equation}
\item[2)]If the above verifications are correct, the TDC will recover the shared key:
\[
 s=K_{B}\oplus K_{D}\oplus K_{i}.
\]
\end{enumerate}

So all participants collaborate together to recover the shared key $s$.

\section{Security analysis}
The security of this protocol is proved in this section. Next, it will be analyzed through the external and internal attacks.
\subsection{External attack}
\subsubsection{Intercept-and-resend attack}
The first attack strategy adopted by the eavesdropper Eve is the intercept-and-resend attack. It can be achieved by intercepting the subsequences sent by the sender, and resends the subsequences forged by Eve to the receiver. It mainly analyzes the following two situations:
\begin{enumerate}
\item[1)]The subsequences of Eve retransmission contain forged decoy particles.
\item[2)]When Eve resends the subsequences, the forged information particles is included.
\end{enumerate}

For the first case, the decoy particles $\{|0\rangle,|1\rangle,|+\rangle,|-\rangle\}$ are randomly selected and inserted into the subsequences, and the sender will not announce the positions of the decoy particles and the corresponding measurement bases until the receiver receives the subsequences.

And if Eve wants to achieve the intercept-and-resend attack, she needs to know the information of the decoy particles before the eavesdropping detections, otherwise she will be found. Thus, this attack is carried out by Eve without knowing any information about the decoy particles, then the probability that Alice and the participants find this attack is $1-(\frac{3}{4})^{m}$ \cite{reff} ($m$ is the number of the decoy particles). When $m$ is large enough, the probability of eavesdropping being discovered approaches to 1.

And for the second case, if Eve resends the wrong information particles, it can't satisfy the equations (\ref{eq1}), (\ref{eq2}) and will be discoverd by the TDC. Then the protocol will be terminated and the wrong shared key can't be recovered.

Therefore, it can be proved that the intercept-and-resend attack is invalid for this protocol.

\subsubsection{Entangle-and-measure attack}

The second attack strategy adopted by the eavesdropper Eve is the entangle-and-measure attack. Assuming that Eve prepares an auxiliary quantum state $|E\rangle$, she entangles the auxiliary particle on the transmitted particle by performing the unitary operation $U_{E}$, and steals the secret information by measuring the auxiliary particle.

The unitary operation $U_{E}$ is defined as follows:
\[
U_{E}|0\rangle|E\rangle=a|0\rangle|E_{00}\rangle+b|1\rangle|E_{01}\rangle.
\]
\[
U_{E}|1\rangle|E\rangle=c|0\rangle|E_{10}\rangle+d|1\rangle|E_{11}\rangle.
\]
where $|a|^{2}+|b|^{2}=1$, $|c|^{2}+|d|^{2}=1$. Since decoy particles are contained in the protocol, the unitary operation $U_{E}$ must satisfy the following conditions:
\begin{equation}
\begin{aligned}
U_{E}\left(|0\rangle|E\rangle\right)&=a|0\rangle|E_{00}\rangle.\nonumber\\
U_{E}\left(|1\rangle|E\rangle\right)&=d|1\rangle|E_{11}\rangle.\nonumber\\
U_{E}\left(|+\rangle|E\rangle\right)&=\frac{1}{2}|+\rangle\left(a|E_{00}\rangle+b|E_{01}\rangle+c|E_{10}\rangle+d|E_{11}\rangle\right)\nonumber\\
&+\frac{1}{2}|-\rangle\left(a|E_{00}\rangle-b|E_{01}\rangle+c|E_{10}\rangle-d|E_{11}\rangle\right).\
\end{aligned}
\end{equation}
\begin{equation}
\begin{aligned}
U_{E}\left(|-\rangle|E\rangle\right)&=\frac{1}{2}|+\rangle\left(a|E_{00}\rangle+b|E_{01}\rangle-c|E_{10}\rangle-d|E_{11}\rangle\right)\nonumber\\
&+\frac{1}{2}|-\rangle\left(a|E_{00}\rangle-b|E_{01}\rangle-c|E_{10}\rangle+d|E_{11}\rangle\right).\
\end{aligned}
\end{equation}

In order to avoid the increase of the error rate, the unitary operation $U_{E}$ must meet the following conditions when Eve introduces the auxiliary particle:
\[
a|E_{00}\rangle+c|E_{10}\rangle=b|E_{01}\rangle+d|E_{11}\rangle.
\]
\[
a|E_{00}\rangle-c|E_{10}\rangle=-b|E_{01}\rangle+d|E_{11}\rangle.
\]

It is easy to get $a=d=1$, $b=c=0$ and $|E_{00}\rangle=|E_{11}\rangle$. So we have the following equations:
\[
U_{E}\left(|0\rangle|E\rangle\right)=|0\rangle|E_{00}\rangle.
\]
\[
U_{E}\left(|1\rangle|E\rangle\right)=|1\rangle|E_{11}\rangle.
\]

Therefore, no matter what the useful state is, Eve can only get the same information from auxiliary particles. Thus, the entangle-and-measure attack cannot succeed in the protocol.

\subsubsection{Trojan horse attack}

The third attack strategy adopted by the eavesdropper Eve is Trojan horse attack. The photons used in the protocol may be insecure to against the two types of Trojan horse attack, that is, the delay-photon attack \cite{refg} and the invisible photon attack \cite{refh,refi}.

At first, in order to prevent the delay-photon attack, the participants can extract a part of the photons and split each particle by the photon number splitter (PNS). Then they use the corresponding measurement bases to measure the photons. If the multi-photon rate is much higher than expected, the PNS will find this attack.

Next, a wavelength optical device that filters out the invisible photons can be installed by the participants to prevent the invisible photon attack. The optical device allows the operational photons to enter, while the invisible photons that belong to Eve will be eliminated.

Therefore, this protocol can completely resist Trojan horse attack.

\subsection{Internal attack}
\subsubsection{Distributor attack}
In this protocol, the distributor Alice needs to perform two key operations:
\begin{enumerate}
\item[1)]Alice needs to prepare six-qubit cluster states $|C\rangle$ correctly.
\item[2)]When all participants complete eavesdropping detections and the X operations, Alice need use the cluster states to correctly measure the subsequences that she obtains, and calculates the hash function values corresponding to the measurement results.
\end{enumerate}

For the first operation, Alice needs to prepare six-qubit cluster states and send the subsequence $S_{3}$, $S_{4}$ ($S_{5}$, $S_{6}$) to ${\rm P}_{i-1}$ (${\rm P}_{i+1}$). The attack strategy that the dishonest distributor may adopt is to prepare the wrong $|C\rangle$ to destroy the protocol. Nevertheless, ${\rm P}_{i-1}$ and ${\rm P}_{i+1}$ need to verify the cluster states prepared by Alice in the implementation of the protocol. If the related properties of  six-qubit cluster states are satisfied, Alice prepares them correctly. Otherwise, Alice is dishonest. And this protocol will terminate.

For the second operation, if the dishonest distributor  attempts to make the recovered shared key wrong, she will be detected by the TDC. Because the TDC can recover the shared key when the equations (\ref{eq1}),(\ref{eq2}),(\ref{eq3}),(\ref{eq4}) and (\ref{eq5}) are all satisfied, that is, $K_{B}$, $K_{D}$, $H(B)$ and $H(D)$ sent by Alice are required to be correct. Thus, the recovered shared key  must be correct.

As a result, the distributor attack could not succeed.

\subsubsection{Participant attack}

In this protocol, the attack strategies that the dishonest participant may adopt are:
\begin{enumerate}
\item[1)]Through the conspiracy attack, participants steal secret information in advance in the process of protocol execution.
\item[2)]The participants send the wrong information to realize the forgery attack, which leads to the error of the recovered shared key.
\end{enumerate}

For the first attack strategy, the only way the participants can take is ${\rm P}_{i}$, ${\rm P}_{i-\frac{n-1}{2}}$ and ${\rm P}_{i+\frac{n-1}{2}}$ conspire to attack. However, $S_{1}^{'}$ and $S_{2}^{'}$ is preserved by Alice, so they cannot recover $s$ in advance.

For the second strategy, the participant ${\rm P}_{j}$ $(j=1,2,$

\noindent $\cdots,n)$ has sent the hash function value $H(K_{i})$ with identity to the TDC in the preparation phase. Before recovering the shared key $s$, the TDC verifies whether the equations (\ref{eq1}), (\ref{eq2}) and (\ref{eq5}) hold. If the verification fails, the protocol will be terminated to avoid recovering the wrong shared key $s$.

So, the participant attack cannot be successful in this protocol.
%
%
%
%
\section{Performance analysis}
In this section, the correctness of the protocol is discussed, and our protocol is compared with three existing MPQKA schemes.

\subsection{Correctness analysis}
\subsubsection{The correctness analysis of the protocol}
The correctness of this protocol will be analyzed next.

\begin{figure}
\resizebox{0.5\textwidth}{!}{%
  \includegraphics{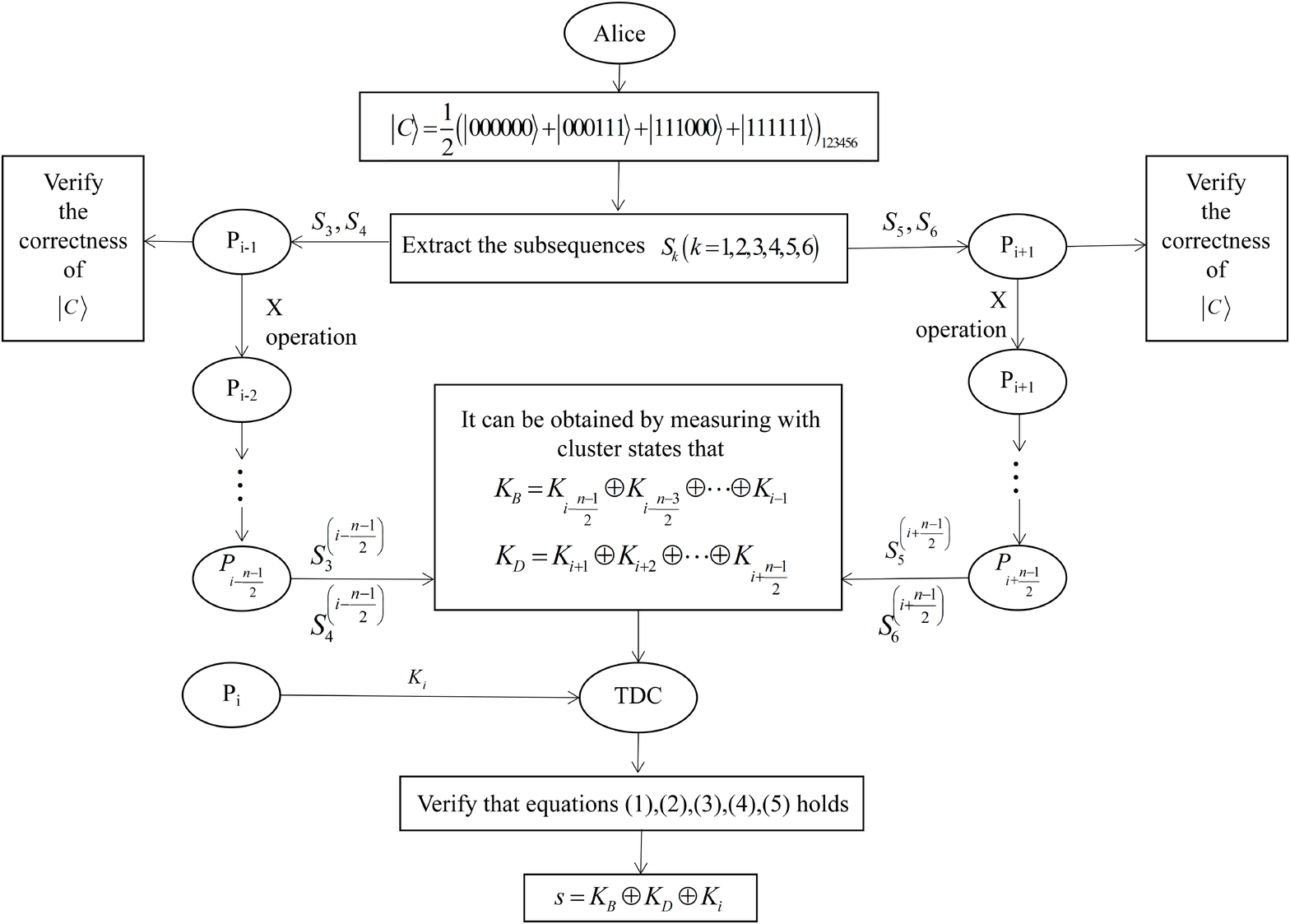}
}
\caption{Protocol simplified flow chart}
\label{fig:1}       
\end{figure}

In order to implement verifiable multi-party quantum key agreement protocol, the participants  ${\rm P}_{j} (j=1,2,\cdots,n)$ needs to generate secret key $K_{j}$  randomly and send the corresponding hash function values $H(j)$  to the TDC. The flow chart of this protocol is shown in Fig.~\ref{fig:1}.

We can seen from Fig.~\ref{fig:1} that the shared key can be recovered only when the equations (\ref{eq1}), (\ref{eq2}), (\ref{eq3}), (\ref{eq4}) and (\ref{eq5}) hold, and

\begin{equation}
\begin{aligned}
s &= K_{B}\oplus K_{D}\oplus K_{i}\nonumber\\
&=K_{i-\frac{n-1}{2}}\oplus K_{i-\frac{n-3}{2}}\oplus \cdots \oplus K_{i-1}\oplus K_{i+1} \nonumber\\
&\cdots \oplus K_{i+\frac{n-3}{2}}\cdots \oplus K_{i+\frac{n-1}{2}}\oplus  K_{i}\nonumber\\
&=K_{1}\oplus K_{2}\cdots \oplus K_{n}.
\end{aligned}
\end{equation}

Therefore, it can be proved that our protocol is correct.
\subsubsection{The example of the protocol}
Next, a simple example is given to make our protocol easy to understand.
We can take $n=5$, $m=3$, $l=2$. Namely there is an distributor Alice and five participants ${\rm P}_{i} (i=1,2,\cdots,5)$, where  $x_{1}=1,x_{2}=2,\cdots,x_{5}=5$.

Then the participant ${\rm P}_{i} (i=1,2,\cdots,5)$  randomly generates secret key  $K_{i}$. We can take  $K_{1}=(10, 11, 10)$, $K_{2}=(00, 01, 01)$, $K_{3}=(11, 01, 00)$, $K_{4}=(11, 10, 11)$, $K_{5}=(00, 10, 01)$. And the corresponding hash function value $H(i)$  with identity is calculated and sent to TDC.

\begin{enumerate}
\item[1)]Distributor operation
\begin{enumerate}
\item[i)]Alice prepares  $m+l=5$ six-qubit cluster states $|C\rangle$. Then the  $k$-th qubit is extracted from each cluster state to form the subsequences  $S_{k}(k=1,2,\cdots,6)$, where  $S_{1}, S_{2}$ is preserved by Alice, and  $S^{*}_{3}, S^{*}_{4}, S^{*}_{5}, S^{*}_{6}$  is obtained by inserting decoy particles into  $S_{3}, S_{4}, $

    $ S_{5},S_{6}$.
\item[ii)]Alice sends $S^{*}_{3}$  and  $S^{*}_{4}$ ($S^{*}_{5}$  and  $S^{*}_{6}$) to  ${\rm P}_{2}$ (${\rm P}_{4}$). After they have been ensured to receive the subsequences, Alice publishes the location of the decoy particles and the measurement bases she used, ${\rm P}_{2}$  and ${\rm P}_{4}$  measure the results and inform Alice. If the results are correct, the protocol continues. ${\rm P}_{2}$ (${\rm P}_{4}$)  recovers   $S_{3}$ and $S_{4}$  ($S_{5}$ and $S_{6}$).
\end{enumerate}
\item[2)]Authentication of distributor identity

${\rm P}_{2}$ and ${\rm P}_{4}$  specify the particle position  $l={4,5}$, and Alice is required to use the X-basis to measure.

Alice and ${\rm P}_{2}$, ${\rm P}_{4}$  use the X-basis to measure the particles where the positions are 4 and 5 in $S_{1}$, $S_{2}$; $S_{3}$, $S_{4}$; $S_{5}$, $S_{6}$  respectively.

When Alice declares that the measurement result is 0, the measurement results of  ${\rm P}_{2}$ and ${\rm P}_{4}$ should be either 0 simultaneously or 1 simultaneously.

When Alice declares that the measurement result is marked as 1, the measurement result of ${\rm P}_{2}$ and ${\rm P}_{4}$ must satisfy one is 0 and the other is 1.

If it is verified that the above requirements are met, Alice has prepared  $|C\rangle$ correctly. Alice, ${\rm P}_{2}$ and ${\rm P}_{4}$ eliminate the two particles in the corresponding position in the subsequences $S_{k}(k=1,2,\cdots,6)$  to form the new subsequence $S^{'}_{k}(k=1,2,\cdots,6)$ . And we move on to the next step.
\item[3)]Participants operation
\begin{enumerate}
\item[i)]After the verification of the cluster states prepared by Alice, ${\rm P}_{2}$  gets $S^{'}_{3}$  and  $S^{'}_{4}$,  ${\rm P}_{4}$  gets $S^{'}_{5}$  and  $S^{'}_{6}$.

 ${\rm P}_{2}$ uses   $K_{2}=(00, 01, 01)$ to operate the X operation on $S^{'}_{3}$  and  $S^{'}_{4}$ to obtain $S^{(2)}_{3}$  and  $S^{(2)}_{4}$, which inserts into the decoy particles to obtain and send the subsequences $S^{(2*)}_{3}$, $S^{(2*)}_{4}$ to  ${\rm P}_{1}$.

Similarly,  $K_{4}=(11, 10, 11)$ is used by ${\rm P}_{4}$  to operate the X operation on  $S^{'}_{5}$  and  $S^{'}_{6}$ to obtain  $S^{(4)}_{5}$  and  $S^{(4)}_{6}$, which inserts into the decoy particles to obtain and send the subsequences $S^{(4*)}_{5}$, $S^{(4*)}_{6}$   to  ${\rm P}_{5}$.
After they are ensure to receive the subsequence,  ${\rm P}_{2}$ and ${\rm P}_{4}$  announce the position of the decoy particles and the measurement basis they used, ${\rm P}_{1}$ and ${\rm P}_{5}$  measure the results and inform ${\rm P}_{2}$ and ${\rm P}_{4}$. If the results are correct, the protocol continues.
\item[ii)]With the same method as i),  ${\rm P}_{1}$ uses $K_{1}=(10, 11, $

\noindent $10)$  to get $S^{(1*)}_{3}$, $S^{(1*)}_{4}$, and  ${\rm P}_{5}$ utilizes  $K_{5}=(00, 10, $

\noindent $01)$ to get $S^{(5*)}_{5}$ and $S^{(5*)}_{6}$. Then  ${\rm P}_{1}$ and ${\rm P}_{5}$  send the subsequences to Alice. After confirming that Alice has received the subsequences, ${\rm P}_{1}$ and ${\rm P}_{5}$ publish the positions of the decoy particles and the measurement basis they used. Next, Alice gets the result and informs ${\rm P}_{1}$ and ${\rm P}_{5}$. If there is no wrong, the protocol continues.
\end{enumerate}
\item[4)]Measurement

Alice uses $S^{'}_{1}$, $S^{'}_{2}$, $S^{(1)}_{3}$, $S^{(1)}_{4}$, $S^{(5)}_{5}$, $S^{(5)}_{6}$ to measures the $j$-th ($j=1,2,3$) particle with the six-qubit cluster states.

    When $j=1$,

    $\frac{1}{2}(|001011\rangle+|001100\rangle+|110011\rangle-|110100\rangle)=|C_{12}\rangle$.

    When $j=2$,

    $\frac{1}{2}(|001000\rangle+|001111\rangle+|110000\rangle-|110111\rangle)=|C_{9}\rangle$.

    When $j=3$,

    $\frac{1}{2}(|001110\rangle+|001001\rangle+|110110\rangle-|110001\rangle)=|C_{15}\rangle$.

    So Alice can get  $K_{B}=(10, 10, 11)$, $K_{D}=(11, 00, 10)$, and calculate the corresponding values of hash function  $H(B), H(D)$.
\item[5)]TDC operation

Alice sends $K_{B}$, $K_{D}$, $H(B), H(D)$ to TDC. At the same time,  $K_{3}=(11, 01, 00)$ are sent to TDC by  ${\rm P}_{3}$.
\begin{enumerate}
\item[i)]The following verifications are performed in TDC.
\[
H(K_{1})H(K_{2})=H(B).
\]
\[
H(K_{4})H(K_{5})=H(D).
\]
\[
H(B)=H(K_{B}).
\]
\[
H(D)=H(K_{D}).
\]
\[
H(3)=H(K_{3}).
\]
\item[ii)]If the above equations are verified correctly, all participants can successfully recover the shared key  $s=K_{B}\oplus K_{D}\oplus K_{3}=(10, 11, 01)$ simultaneously.
\end{enumerate}
\end{enumerate}

\subsection{Comparation analysis}
In this section, we will compare our protocol with three existent MPQKA protocols.

Ever since the QKA protocol was proposed in 2004, most of QKA protocols have the problem that only one participant can determine the shared key alone. Therefore, the fairness can't be achieved. Next, we compare our protocol with three MPQKA protocols that can achieve fairness, namely Liu et al.' s protocol \cite{refV}, Xu et al. 's protocol \cite{refW},  and Liu and Liang's protocol \cite{refb}. For simplicity, we call them  LG protocol, XW protocol and LL protocol.

For QKA protocol, quantum efficiency \cite{refj} is defined as follows:
\[
\eta=\frac{c}{q+b}.
\]
where $c$ is the length of the final recovered shared key, $q$ represents the total number of bits used in the quantum channel, and the number of bits used in the classic channel is denoted by $b$.

In our protocol, $2m$ is the number of bits of the shared key. We know that  Alice prepares $m+l$ six-qubit cluster states  for ${\rm P}_{i-1}$ and ${\rm P}_{i+1}$ during the distributor operation, and she needs $6(m+l)$ decoy particles. However, the subsequences $S_{1}$ and $S_{2}$ are preserved by Alice, only four subsequences of six-qubit cluster states and $4m$ decoy particles are needed in the participant operation phase. Besides,  $6m$ particles are sent to the TDC by Alice and ${\rm P}_{i}$ by the classical channel. According to this, the efficiency of our protocol can be calculated as:
\begin{equation}
\begin{aligned}
\eta&=\frac{2m}{\{6(m+l)+6(m+l)\}+\left(4m+4m\right)\cdot\frac{N-1}{2}+6m}\nonumber\\
&\approx\frac{1}{2N}.\
\end{aligned}
\end{equation}
where $l$ is the number of particles used to verify Alice's identity. For convenience, we make $l=m$. Therefore, the efficiency of our protocol is approximately equal to $\frac{1}{2N}$.

%
%
%
%

\begin{table}
\caption{Comparison between the existent protocols and our protocol}
\resizebox{0.5\textwidth}{!}{%
  \includegraphics{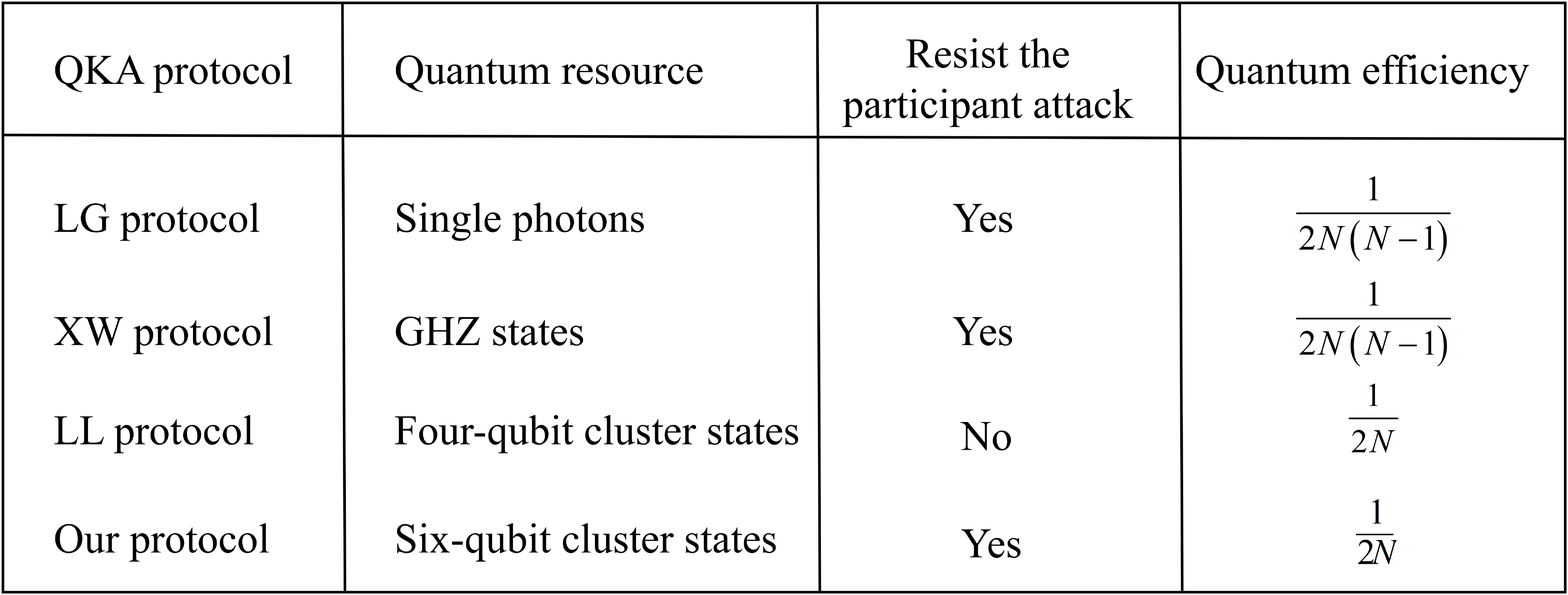}
}
\label{Table:2}       
\end{table}

Then as shown in Table~\ref{Table:2}, the existing three MPQKA protocols are compared with our protocol.

 It can be seen from Table~\ref{Table:2} that the efficiency of our protocol is increased compared with XW protocol and LG protocol. Besides, our protocol can resist the participant attacks that LL protocol cannot. Therefore, our protocol is effective.

\section{Summary}
Quantum key agreement should satisfy four security features, namely security, correctness, fairness and privacy. We propose a MPQKA protocol based on six-qubit cluster states in this paper. A distributor that each operation can be verified is introduced to resist the internal fraudsters and the external eavesdroppers attack. And the protocol ensures the correctness and simultaneous of the shared key by using the homomorphic hash function and the TDC. Besides, all participants collaborate together to recover the shared key, which guarantee the fairness and the privacy of the protocol.

\subsection*{Acknowledements}
We would like to thank anonymous review for valuable comments. This work is supposed by the National Natural Science Foundation of China under Grant No.11671244.


%
%



\end{document}